# Understanding the Complexity of Terrorist Networks

Philip Vos Fellman, Southern New Hampshire University

*Abstract*— **Complexity science affords a number of novel tools for examining terrorism, particularly network analysis and NK-Boolean fitness landscapes. The following paper explores various aspects of terrorist networks which can be illuminated through applications of non-linear dynamical systems modeling to terrorist network structures. Of particular interest are some of the emergent properties of terrorist networks as typified by the 9-11 hijackers network, properties of centrality, hierarchy and distance, as well as ways in which attempts to disrupt the transmission of information through terrorist networks may be expected to produce greater or lesser levels of fitness in those organizations.**

## I. INTRODUCTION

OPEN source acquisition of information regarding terrorist networks offers a surprising array of data which social network analysis, dynamic network analysis (DNA) and NK-Boolean fitness landscape modeling can transform into potent tools for mapping covert, dynamic terrorist networks and for providing early stage tools for the interdiction of terrorist activities. Network theory had its beginnings in the work of 18th century mathematician Leonard Euler, with Euler's best know example being the famous "7 bridges of Königsberg" problem. In this context, the problem was found to be insoluble because it failed to meet the specifications of an Euler path. An Euler path is a continuous path that passes through every arc once and only once. In its modern formulation, finding Euler paths and Hamiltonian paths on networks involves finding the minimal edge and vertex paths and the measurement of minimal vertex degree. In particular, "A network is Hamiltonian-connected if it contains a Hamiltonian path between every two distinct nodes. In other words, a Hamiltonian-connected network can embed the longest linear array between any two distinct nodes with dilation, congestion, load, and expansion all equal to one." [1]

## II. PRIOR ART

In our earlier work on Modeling Terrorist Networks [2] we focused on the strategic applications of first principles of counter-intelligence and complex systems to provide insights to both practitioners and theorists at what we described as "the mid-range" of decision making. The following description of how I approached the task of structuring our enterprise may be of particular use in the present applied context:

"As we started thinking about how to organize this paper, I was reminded of two statements which powerfully impressed me as a student of National Security Policy at Yale University over two decades ago. Paul Bracken (1984), who had just recently left the Hudson Institute, where had worked a number of years for Hermann Kahn, the "father of modern nuclear strategy" impressed upon us very early on that National security policy, which in those days was primarily focused upon the command and control of nuclear forces, is characterized by *irreducible levels of ambiguity and complexity*.

The mathematical revolution of chaos theory and complexity science has given us powerful new modeling tools that were unthinkable just a generation ago. The pace of technological change has matched the emergence of new sciences in ways which likewise would have been difficult to conceive of in even the relatively recent past. For example, one of the primary purposes of the U.S. 1979 Export Administration Act was to prevent the migration of dual-use technologies like the 32 bit architecture of the Intel 80486 Microprocessor, which could be used as a targeting system for ICBM's.

Today there's more science and technology involved in controlling the scaling dynamics of internet traffic packet delay than there is in designing the navigational system for mid-course correction on any ballistic missile re-entry system. When these kinds of technological developments are combined with the new power of autonomous non-state actors and various persistent vulnerabilities of complex, self-organizing systems [3], the challenges of national security in the 21st century truly take on an entirely different character and require, tools, techniques, resources, models and knowledge which are fundamentally different from their 20th century predecessors.

In the context of this newly emerging dynamics, a proper

---

**Philip V. Fellman**, B.F.A., California Institute of the Arts (1974), M.A.., Cornell (1979), M.B.A. Yale, (1986), M.A. (Cornell, 1988), Ph.D. (Cornell, 1990), is Professor of International Business Strategy at Southern New Hampshire University in Manchester, New Hampshire, U.S.A.

Dr. Fellman has been a frequent contributor on terrorism to a wide variety of conferences and publications. He is an active member of the New England Complex Systems Institute, where he is currently involved with Dr. Yaneer Bar-Yam and Dr. Ali Minai in co-authoring a volume on using complex adaptive systems research to counter terrorism, insurgency, regional and ethnic violence.

approach to modeling terrorist networks and disrupting the flow of information, money, material needs to be found in such a way that at the **mid-range**, various government agencies can efficiently share information, spread and reduce risk (particularly risk to sensitive infrastructure) as well as pro-actively *target and dismantle* the specific components of embedded terrorist networks while deconstructing their dynamics."

In recent years, our work, and several important works of Kathleen Carley's CASOS group and their graduates have focused more narrowly on the isolation of terrorist group components and the use of isolation techniques to disrupt the flow of information, resources, and command decisions through the network. In the same fashion, Serge Galam has used a percolation model from physics to take a novel approach to isolating violent terrorist cells under conditions of insurgency. [4]

At the same time, from a Homeland Security perspective, the traditional problems of following the money, following the terrorists and following the weapons remain as paramount priorities. In this context, we have attempted to relate the tools of complexity science to the first principles of counter-intelligence in the ways in which we apply those tools to the problems of terrorism. In particular, I would like to emphasize that we're not modeling here for the sake of making theoretical breakthroughs, but rather we seek to apply the models to traditional principles of *coverage*, *penetration* and *isolation* (over-compartmentation).

III. MAPPING TERRORIST NETWORKS

A variety of network maps and mapping techniques have emerged in the post 9-11 World. One of the earliest and most influential maps was developed by Valdis Krebs (Krebs, 2001) shortly after 9/11 [5]. While this map contains some inaccuracies, its overall value is demonstrated by the ability to map the degree of centrality as well as network density with respect to the 9/11 hijacker network.

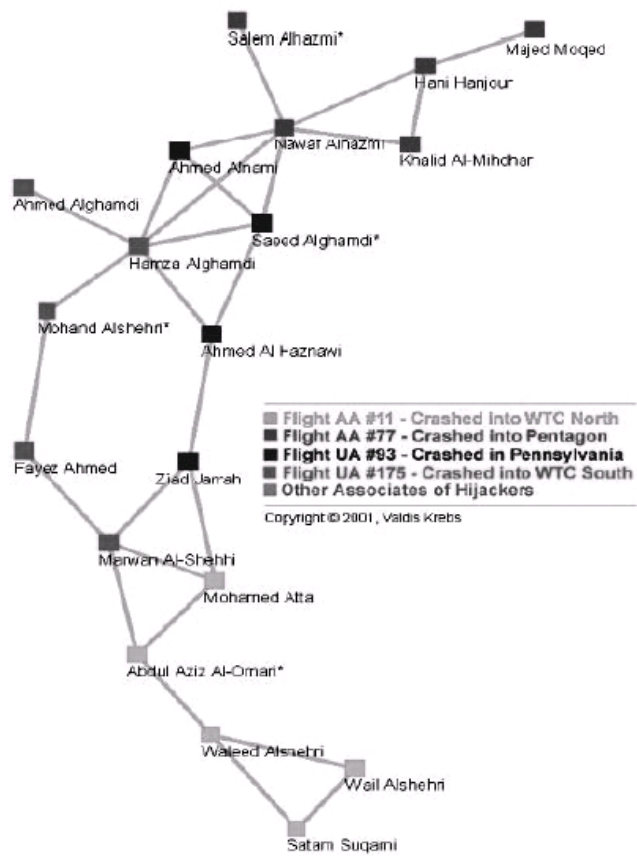

Figure 1: Krebs' 9-11 Hijacker Network Mapping

Using measures of centrality, Krebs' work analyzes the dynamics of the network. In this regard, he also illuminates the centrality measure's sensitivity to changes in nodes and links. In terms of utility as a counter-intelligence tool, the mapping exposes a concentration of links around the pilots, an organizational weakness which could have been used against the hijackers had the mapping been available prior to, rather than after the disaster, suggesting the utility of developing these tools as an ongoing mechanism for combating terrorism.

Once one can measure network density and centrality, a number of new network properties emerge. As Thomas Stewart explains in "Six Degrees of Mohammed Atta" [6] with respect to Krebs' network mapping:

"It is not a complete picture; among other problems, it shows only those links that have been publicly disclosed. Still, it's possible to make some interesting inferences. First, the greatest number of lines lead to Atta, who scores highest on all three measures, with Al-Shehhi, who is second in both activity and closeness, close behind. However, Nawaf Alhazmi, one of the American Flight 77 hijackers, is an interesting figure. In Krebs's number crunching, Alhazmi comes in second in betweenness, suggesting that he exercised a lot of control, but fourth in activity and only seventh in closeness. But if you eliminate the thinnest links (which also tend to be the most recent -- phone calls and other connections made just before Sept. 11), **_Alhazmi becomes the most powerful node in the net._** _He is first in both control and access, and second only to Atta in activity. It would be worth exploring the hypothesis that Alhazmi played a large role in planning the attacks, and Atta came to the fore when it was time to carry them out._"

We have explored this and other possibilities elsewhere [2, 10] and find it worth noting that it may be a more general rule that in many terrorist operations planning is carried out by one major network figure while the operational commander is only brought forward at the actual inception of the operation.

## IV. CARLEY'S DYNET MODEL

The most developed version of these tools is probably the Dynamic Network Analysis (DNA) model developed by Carley et al. [17]

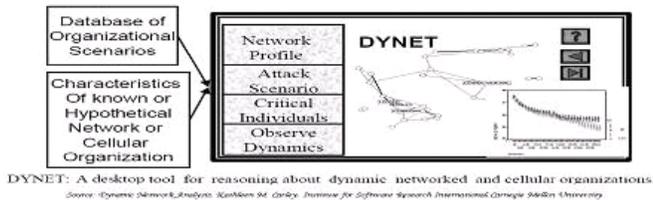

Figure 2: Carley's Dynet Model

As indicated in the figure above, Models of the 9-11 Network, the Al Qaeda network which bombed the U.S. Embassy in Tanzania and other terrorist networks also show a characteristic emergent structure, known in the vernacular as ―the dragon". One element of this structure is relatively large distance [11] between a number of its covert elements [12] which suggests that some elements of the network, particularly certain key nodes [13] may be relatively easy to remove through a process of over-compartmentation or successive isolation, thus rendering the organization incapable of transmitting commands across various echelons.

Carley's Dynet model is also particularly good at distinguishing between cellular and hierarchical networks. The Dynet model is also exceptionally sensitive to the differences in outcome when critical (i.e. leadership) nodes in differing types of networks are removed as shown below [14]:

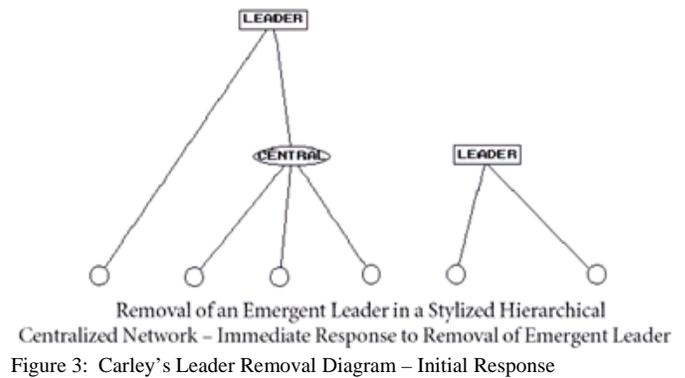

Figure 3: Carley's Leader Removal Diagram – Initial Response

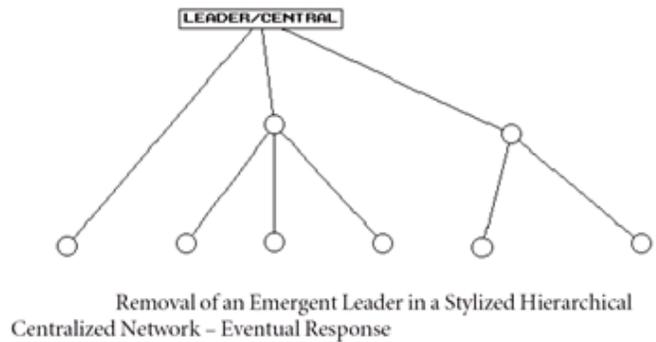

Figure 4: Carley's Leader Removal Diagram – Final State

As Carley explains, "Using DyNet the analyst would be able to see how the networked organization was likely to evolve if left alone, how its performance could be affected by various information warfare and isolation strategies, and how robust these strategies are in the face of varying levels of information assurance.

Another use of Carley's model is one which is quite different than the physical removal of terrorist leaders, or, indeed, any traditional form of "node clipping". One of the most profound implications of the Dynet model is that rather than concentrating on removing terrorist leaders at various levels, a policy of information isolation may better serve to significantly degrade the functioning of terrorist organizations [17]:

"As the world becomes more informated it is expected that technology will of enable greater adaptivity. In general, the idea is that information technology, such as databases, can prevent information loss and facilitate the rapid training of recruits. Preliminary results suggest that this may not be the case…we see that the value of databases for standard networks will be strongest in the long run when after there has been sufficient attrition; whereas, in cellular networks, in the long run databases may actually serve to exacerbate the destabilization caused by attrition. One reason for this is that in the cellular network the strong divisionalization that enabled adaptivity is broken by the database. The results shown are for the case where the personnel use a mixed information search strategy, attrition is random but sustained, and the database is initially incomplete."

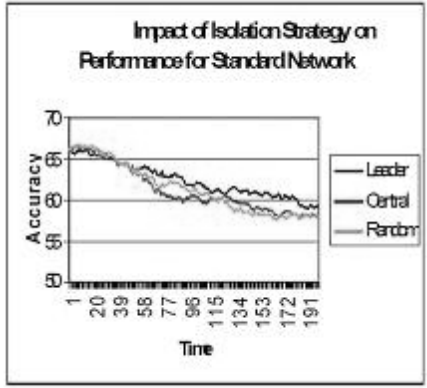

Figure 5: Information Isolation for Standard Network

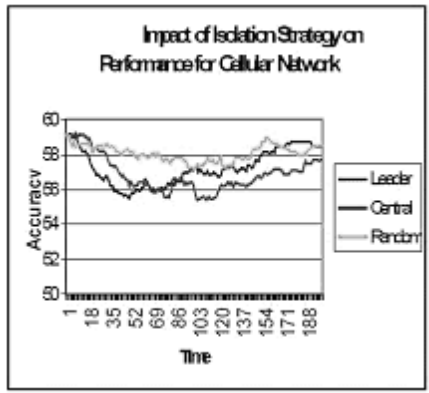

Figure 6: Information Isolation for Cellular Network

As indicated by the above diagrams, DNA provides a number of new insights into what happens when leadership nodes are removed from different kinds of networks and how the process of node removal yields very different kinds of results depending upon whether the network is cohesive or adhesive [7]. Beyond this, as show in figures 5 and 6, DNA shows how isolation strategies (also suggested in Fellman and Wright [2]) yield different results based on the nature of the network involved. Finally it supports our contention that removing terrorist leaders often contributes more to expanding rather than containing the problem of terrorism. In the vast majority of cases, where the networks are cellular in nature, it would be more effective to provide misleading and false data and to isolate the network rather than physically removing leadership targets, particularly when these are "soft targets" and middle or low level leaders [23].

## V. SYMMETRIES AND REDUNDANCIES

Another property of the 9-11 Network is that it demonstrates a high degree of redundancy as illustrated by the following diagram [10].

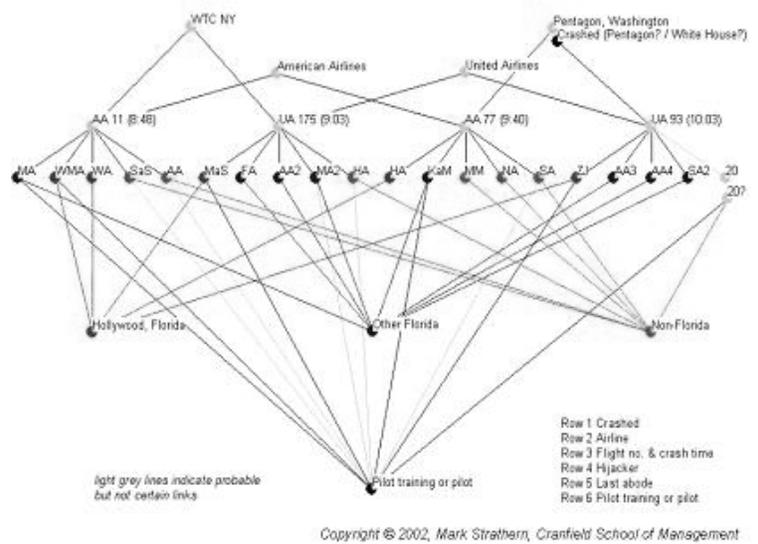

Figure 7: They Symmetries and Redundancies of Terror

This mapping suggests, in turn, that while the network may be highly distributed [11], the redundancies built into it suggest cohesion at a number of levels as well as an hierarchical organization [12]. A number of systems analysis tools [13] have been developed to deal with the problem of covert networks [14] and incomplete information [16]. Again the data presented in the above graphical form suggests that isolation is probably also an easier strategy to perform than node removal because terrorist nodes are specifically constructed to have multiple, redundant nodes for their operational features.

## VI. TERRORIST NETWORKS AND FITNESS LANDSCAPES

The dynamic NK-Boolean fitness landscape models developed by Stuart Kauffman [18, 19] for evolutionary biology have attracted increasing interesting in a number of fields involving social phenomena. In particular, it offers a substantial degree of promise in providing new models for understanding and evaluating the strategic performance[20] of organizations [21].

In 2000, with revisions through 2005, Pankaj Ghemawat of Harvard and Daniel Levinthal of the Wharton School ran an NK-Boolean Dynamic Fitness Landscape simulation, producing over 1,000,000 fitness landscapes with approximately 75 characteristic organizational structures in order to test strategic planning hypothesis about what makes an organization robust.

Levinthal and Ghemawat used a system of adjacency matrices to characterize decisions in hierarchical, distributed (i.e., cellular) and randomly structured organizations. They tested robustness by injecting false data, running histograms and applying replicator dynamics to determine the organization's sensitivity to bad data, in terms of disruption and recovery across ten different policy variables, six different levels of search strategy and the three above specified organizational types.

In 2005 and 2006 we adapted their study to terrorist organizations and drew a number of conclusions from their study, the most significant having to do with organizational search patterns and with the constraint of key variables [23].

When Ghemawat and Levinthal explored the differences between an optimal preset of policy configurations between hierarchical, central and random simulation arrays they found that disrupting lower order policy variables was likely to have little relative effect on the robustness of the organization as compared to the mis-specification of higher order values. This effect was particularly marked in hierarchical organizations, as illustrated below.

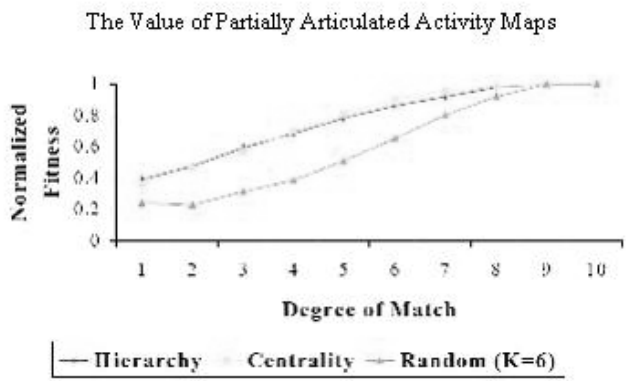

Figure 8: The Results of Mis-Specified Information Values

In dealing with terrorist organizations, especially those such as Jemaah Islamiyah whose organizations are primarily hierarchical in nature [24, 27], the importance of this finding is that it demonstrates the fact that if disinformation is to be useful tactic (or strategy) it must succeeds in influencing one of the key decisional variables. In other words, disinformation, as compared to the informational isolation strategies discussed above, when applied at the local level is unlikely to have any lasting impact on terrorist organizations. This finding also challenges the institutional wisdom [25] of assigning case officers in the field to this type of counter-terrorism operation [26]. In fact, from an operational point of view, the hierarchical nature of terrorist organizations means that there may be something of a mismatch in the entire targeting process. As Ghemawat and Levinthal note, "*Less central variables not only do not constrain, or substantially influence the payoff of many other choices*, but they themselves *are not greatly contingent upon other policy choices*. Being contingent on other policy choices facilitates compensatory shifts in policy variables other than the one that is preset. As a result of the absence of such contingencies, the presetting of lower-order policy choices is more damaging to fitness levels under the centrality structure." (p. 27)

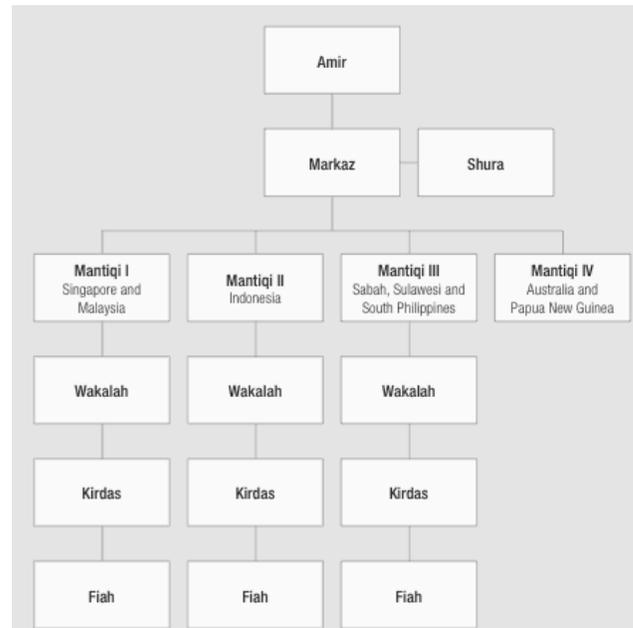

Figure 9: Centrality - The Hierarchical Structure of Jemaah Islamiyah

Another striking feature of this set of simulations concerns how few of the optima with preset mismatches constitute local peaks of the fitness landscape. Given the importance of configurational effects, one might reasonably conjecture that constraining one variable to differ from the global optimum would lead to the selection of a different, non-global, peak in the fitness landscape. However, in the actual simulation runs this result was unexpectedly absent.

Instead, constraining the independent variables to differ from the global optimum had a larger impact than constraining the seemingly more important "contingent" variables. The reason for this is that the presence of contingency allows for the possibility of substituting or compensating changes in policy variables. While tightly linked interaction patterns have generally been viewed as fragile, the equifinality that high levels of interaction engender also allows for a certain robustness, a conclusion which ties in well with Carley's Dynet model.

In contrast, when an autonomous variable is misspecified, that doesn't create negative ramifications elsewhere in the system of policy choices; at the same time, however, there is no opportunity to compensate for the misspecification. These results are illustrated below in Ghemawat and Levinthal's normalized fitness level graph showing what they describe as "The Constraints of History" [22].

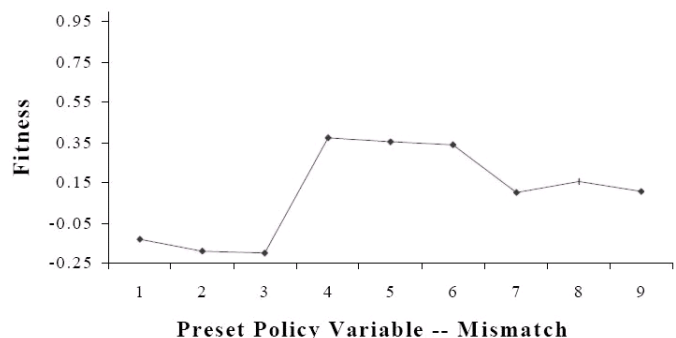
Figure 10: Normalized Fitness Levels and the Constraints of History

As this diagram demonstrates, for disinformation to produce effective results, the policy mismatch needs to occur at the top of the decisional chain and the difference between variable mis-specification is most pronounced with respect to the first three variables. In fact the effects of constraining any of these first three variables outweighs all other efforts combined.

## VII. Conclusion

Terrorist networks are complex. They are typical of the structures encountered in the study of conflict, in that they possess multiple, irreducible levels of complexity and ambiguity. This complexity is compounded by the covert, dynamic nature of terrorist networks where key elements may remain hidden for extended periods of time and the network itself is dynamic. Network analysis, agent-based simulation, and NK-Boolean fitness landscapes offer a number of tools which may be particularly useful in sorting out the complexities of terrorist networks, and in particular, in directing long-run operational and strategic planning so that tactics which appear to offer immediately obvious rewards do not result in long term damage to the organizations fighting terrorism or the societies which they serve.

In a substantive sense, the results which these simulation and dynamical systems modeling tools present suggest that quite literally "sometimes less is more" and that operational objectives might be better directed at isolation rather than removal. Similarly the results described here suggest that "soft targets" are, for the most part, probably not worth pursuing. In other words, if you are not focused on the top problems, then considerations of opportunity cost suggest that it may be better to do nothing than to waste valuable resources on exercises which are doomed to counter-productivity.